# A Modular Measurement Integrity Verification for Transformer Currents with Applications in Hardware-in-the-Loop Digital Twins

Reza Jalilzadeh Hamidi, *Senior Member, IEEE*

*Abstract*—This paper proposes a recursive method for integrity verification of measured transformer currents, which is suitable for the modular development of Hardware-In-the-Loop Digital Twins (HIL DTs). The Differential Equations (DEs) describing transformer transients are relatively complex, requiring the use of numerical DE solvers with small time steps. This implies that replicating transformers with HIL DTs requires a continuous flow of accurate measurement samples with a high sampling rate.

The proposed method utilizes the Adaptive Extended Kalman Filter to estimate the parameters of transformer currents which can be non-sinusoidal during transients such as transformer energization. Then, after evaluating the validity of estimations, the proposed method utilizes the estimated parameters to reconstruct noiseless measurements with desirable sampling rates. The performance of the proposed method is evaluated using EMTP-RV for simulating transformers and Matlab for executing the proposed method. The simulation results demonstrate that the method is able to accurately estimate and closely track the current measurements.

## I. INTRODUCTION

THE Digital Twin (DT) is a promising technology that enables operators to become aware of the conditions of a physical entity by observing its digital replica [1]. With reference to [1]-[8], the structure of DTs can be divided into three major functional blocks, as follows: 1) the measurement integrity verification block that ensures the accuracy and continuity of the input data [9]. 2) The model which retains the mathematical, logical, data-driven, or hybrid representation of the physical entity. 3) The executor (also referred to as solver) that determines the states of the physical entity by solving the model using the input data [10].

Monitoring power transformers is particularly remarkable in light of their frequent application, cost, and critical role in power delivery [2]. The internal states of a transformer (e.g., core flux) can be determined by replicating the transformer during steady and transient (e.g., energization) states. This necessitates the Differential Equations (DEs) describing transformers are solved by numerical DE solvers with small time steps [5], [11] using real-time measurements from the transformer. Therefore, the DT and transformer together form a Hardware-in-the-Loop (HIL) system in that the DT replicates the transformer in a real-time manner. However, the accuracy of numerical DE solvers largely depends on their time steps and the solidness of their input data [12]. Accordingly, the measurement integrity verification block should be responsible for continuously providing accurate and noiseless measurements with an adequate sampling rate.

There are numerous works on the concept and perspectives of DTs. For example, in [1], [13], the general frameworks and applications of DTs are described for predictive maintenance and diagnosis of electric grids. In [14], DTs are suggested for reliability assessment of distribution grids. In [15] and [16], the potential applications of DTs in cyber security of smart grids and energy management are discussed, respectively.

However, there is limited literature available on the technical aspects of DTs for electric apparatus such as power transformers [8]. In [17], data on different aspects of a transformer is provided for enabling other researchers to develop DTs for transformers. In [4], a DT is proposed that uses sensor data on dissolved gases in transformer oil, oil dielectric losses, insulation resistance, and core grounding current to create an index for health monitoring of transformers. In [18], a DT is described that uses multi-physics principles to estimate the temperature of hot spots in oil-immersed transformers and forecast their remaining operational life. In [19], a DT is introduced that uses oil tank vibrations to diagnose failures in transformer windings. In [2], an HIL DT is detailed for monitoring single-phase or three-phase transformer banks in distribution grids. It receives time-domain voltage and current measurements from the transformer LV-side to calculate the HV-side voltages and currents. In [20], an HIL DT is proposed that receives voltage and current measurements from either side of a transformer and calculates the other side parameters. Then it compares the calculated values and measurements to detect internal failures.

In the aforementioned technical works, it is assumed that the measurements meet the required accuracy, continuity, and sampling rate. The uncertainties in measurements are however addressed in the following works. In [7], the overall structure of DTs and the effects of uncertainties in measurements and models are discussed. The use of the Weighted Least Squares (WLS) and Extended Kalman Filter (EKF) is tested for model and measurement correction. However, this work utilizes Phasor Measurement Unit (PMU) measurements which is improper for DTs modeling transformer transients. In [5], an HIL DT is developed for core-type three-phase transformers that estimates the voltages and currents of each side of the

Reza Jalilzadeh Hamidi is with the ECE Dept. of Georgia Southern University (Email: reza.j.hamidi@gmail.com).

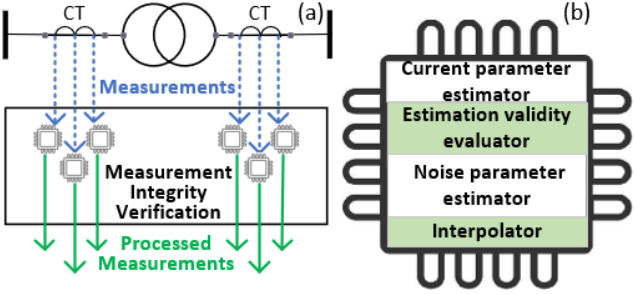

Fig. 1. (a) High-level overview of the proposed method. (b) The modules that are loaded in each processing unit to be rendered.

transformer using the other side measurements. It also estimates the core flux and core losses. In both [5], the measurement integrity is addressed using the Discrete Kalman Filter (DKF) only during the steady-state situation neglecting transients.

With respect to the mentioned requirements for measurement integrity, this paper proposes a computationally effective method for the development of modular measurement integrity verification appropriate for HIL DTs replicating single-phase and core-type three-phase transformers. The modular architecture of the proposed method enables DT users to gradually extend the DT based on their priorities.

The proposed method employs the Adaptive Extended Kalman Filter (AEKF) to recursively estimate the parameters of transformer currents during both steady-state and transient situations. Then, it evaluates the validity of the estimations, and if they are valid, the proposed method reconstructs the measurements with sampling rates suitable for numerical DE solvers.

It should be noted that high sampling rate measurements (e.g., 200 k samples/second per measurement) significantly increase communication traffic. Therefore, it is preferable to transfer measurement streams with lower sampling rates via communications. Then, measurements are resampled locally at the HIL DT at an adequately high rate. The advantages of the proposed method are as follows: 1) it is able to estimate the parameters of measured currents even when the transformer is heavily saturated. 2) It is computationally effective owing to its recursive mathematical formulation which is suitable for implementation in industrial computers. 3) The proposed method adaptively modifies itself with respect to measurement noises which increases its accuracy and trackability.

The rest of this paper is organized as follows: in Section II, the methodology of the proposed method is described. In Section III, test cases are presented and the results are discussed. Finally, the conclusion is presented.

## II. METHODOLOGY

Fig. 1(a) presents a high-level overview of the proposed method, in that each current measuring device sends one stream of samples to one of the parallel processing units that the proposed measurement integrity verification block hosts. Therefore, if the DT user requires to use new measurements, additional parallel processing units can be added to the DT for processing them. Each processing unit is also programmed to run four modules as shown in Fig. 1(b). These modules can be removed based on the user's needs. For example, the noise parameter estimator can be removed based on the user's discretion.

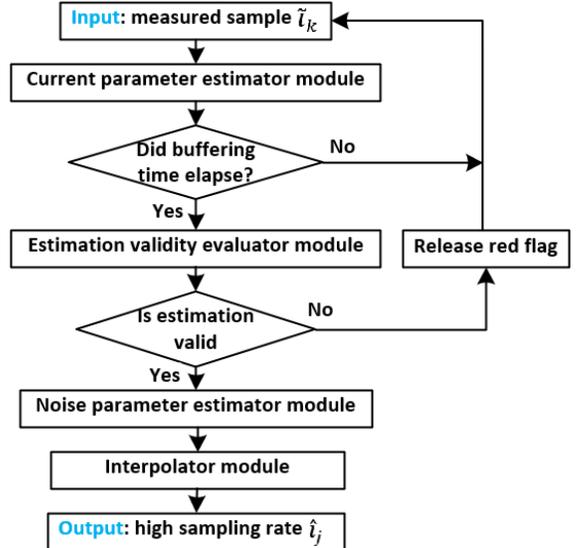

Fig. 2. Flowchart of the proposed method.

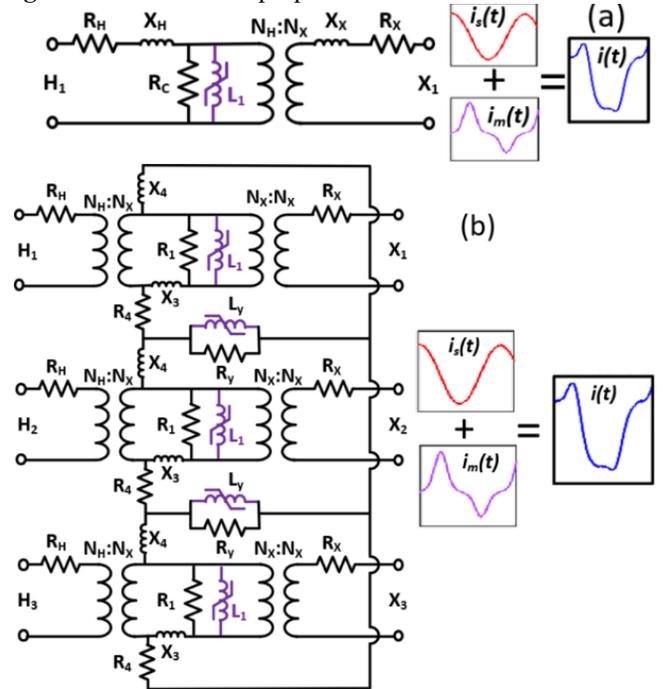

Fig. 3. (a) The Steinmetz model of single-phase transformers. (b) The hybrid model of three-phase transformers. The colored inductors cause the nonlinear component of the transformer current.

As Fig. 2 shows, each measured sample is first sent to the current parameter estimator that removes noises and estimates the parameters of the measured current. After a short buffering time for estimations to become stable, the estimation results are sent to the estimation validity evaluator that evaluates the validity of the estimations. Then, if the estimations are valid, the noise parameter estimator finds the noise parameters. Finally, the interpolator module reconstructs the measurement at a sufficiently high sampling rate. The details of these four modules are explained in the following subsections.

*A. Current Parameter Estimator*

Referring to Fig. 2, the first module processing a measured sample is the current parameter estimator. With reference to Figs. 3(a) and (b), a transformer current comprises two components as

$$i(t) = i_m(t) + i_s(t) \quad (1)$$

where $i(t)$ [A] is the transformer current, $i_m(t)$ [A] is the magnetization current which is often non-sinusoidal due to the core saturation, and $i_s(t)$ [A] is the sinusoidal component of the transformer current. In single-phase transformers (Fig. 3(a)), one nonlinear inductor forms $i_m$, while in core-type three-phase transformers (Fig. 3(b)), several nonlinear inductors form $i_m$.

The magnetization current $i_m$ is proportional to the total flux linkage ($\lambda$) and can be approximated with a two-term polynomial as follows [21], [22]

$$i_m(\lambda(t)) \approx \beta_1 \lambda(t) + \beta_2 \lambda^n(t) \quad (2)$$

where $i_m(\lambda(t))$ [A] is the magnetization current, $\beta_1$, $\beta_2$, and $n$ are device-specific parameters of the core that are explained in [21], [22], and $\lambda(t)$ [Wb] is the core's total flux linkage at any given time $t$. The sinusoidal component of $i(t)$ in (1) is

$$i_s(t) = i_s^p \sin(\omega_0 t + \Theta) \quad (3)$$

where $i_s^p$ [A] is the peak value of the sinusoidal component, $\omega_0$ [rad/s] is the power system angular frequency, and $\Theta$ [rad] is the initial phase of the sinusoidal component.

The measurements are noise-contaminated, necessitating the use of an estimation method to remove the noise. Several methods for measurement estimation are compared, first. Then the selection of the AEKF is rationalized. Per the IEEE Standard 1459-2010 [23], Fourier-based methods have limited trackability, and those are proper for stationary measurements [23], [24]. Although the trackability is improved in the short-time Fourier transform, it causes aliasing, time-window effect, picket fence, and bandwidth trade-off issues [25]. The Savitzky-Golay filter is capable of canceling noises and determining the sequence components of electrical measurements. However, its application in estimation of measurement parameters needed for reconstructing the measurement is limited [26]. The Prony [27], matrix pencil [28], and short-time matrix pencil [29] methods are efficient in noise cancellation. However, they are window-based, and the window size impacts their performance. Moreover, these methods demand complex matrix algebra on large matrices that is challenging to be implemented in industrial computers for real-time applications. The Discrete Wavelet Transform (DWT) offers promising trackability, but the selection of the appropriate mother wavelet and scale influences its accuracy [25]. The Kalman Filter (KF) has been in use for measuring power system transients owing to its high trackability and accuracy [24], [30], [31]. Since transformer inrush currents are nonlinear and the noises in power girds are inconsistent [32], the AEKF is used in the proposed method.

Since the proposed method is based on the AEKF, observer and process dynamic equations should be defined. The observer equation is based on the transformer current. Thus, its constituents (i.e., $i_m$ and $i_s$) are first formulated. $i_m$ is a function of the core's total flux linkage which is in the form of [33], [34]

$$\lambda(t) = \lambda_m \sin(\omega_0 t + \alpha - \phi) + (\lambda_r - \lambda_m \sin(\alpha - \phi))e^{-t/\tau} \quad (4)$$

where $\lambda_m$ [Wb] is the steady-state amplitude of the flux linkage, $\alpha$ [rad] and $\phi$ [rad] are respectively the phase and impedance angle of the source supplying the transformer, $\lambda_r$ [Wb] is the residual flux, and $\tau$ [s] is the decay time constant.

As measuring the residual flux linkage ($\lambda_r$) is practically challenging, and considering that the general pattern of flux linkage variation is required for the proposed method, (4) can be written in a simplified and discrete-time form as

$$\lambda_k = \lambda_m \sin(\omega_0 k T_s + \theta) + \lambda_0 e^{-kT_s/\tau} \quad (5)$$

where $\theta = \alpha - \phi$ [rad] is the phase angle of the steady-state flux linkage, $\lambda_0 = \lambda_r - \lambda_m \sin(\alpha - \phi)$ [Wb] is the initial value of the decaying term, $k$ is the sample number, and $T_s$ [s] is the sampling period. Therefore, referring to Fig. 3(a) and (b), for single-phase transformers, $i_m$ is caused by the only inductor $L_1$, while in the case of core-type transformers, $i_m$ is proportional to a combination of currents flowing in all $L_1$s and $L_y$s. Since the saturation curves of $L_1$s and $L_y$s are different [34], two polynomials with different parameters are employed to approximate $i_m$ with the rationale provided in Appendix A.

$$i_m(\Lambda(t), \Lambda'(t)) \approx \beta_1 \Lambda(t) + \beta_2 \Lambda^n(t) \\ + \beta'_1 \Lambda'(t) + \beta'_2 {\Lambda'}^{n'}(t) \quad (6)$$

where $\Lambda$ and $\Lambda'$ are sinusoidal flux linkages representing a combination of limb and yoke fluxes, respectively. The terms with no prime (.') suffice for single-phase transformers containing only $L_1$, while terms with prime are necessary for core-type transformers to take yokes into account. Both $\Lambda$ and $\Lambda'$ in yokes follow the pattern given in (5) as explained in Appendix A, which can be expanded as

$$\begin{cases} \Lambda_k = \Lambda_m \cos(\theta) \sin(\omega_0 k T_s) + \\ \quad \Lambda_m \sin(\theta) \cos(\omega_0 k T_s) + \Lambda_0 e^{-kT_s/\tau} \\ \Lambda'_k = \Lambda'_m \cos(\theta') \sin(\omega_0 k T_s) + \\ \quad \Lambda'_m \sin(\theta') \cos(\omega_0 k T_s) + \Lambda'_0 e^{-kT_s/\tau'} \end{cases} \quad (7)$$

Accordingly, the following states and coefficients can be defined:

$$\begin{cases} \Lambda_{d_k} := \Lambda_m \cos(\theta), & a_{d_k} := \sin(\omega_0 k T_s) \\ \Lambda_{q_k} := \Lambda_m \sin(\theta), & a_{q_k} := \cos(\omega_0 k T_s) \\ \Lambda_{0_k} := \Lambda_0 e^{-kT_s/\tau}, & a_{0_k} := 1 \\ \Lambda'_{d_k} := \Lambda'_m \cos(\theta'), & a'_{d_k} := \sin(\omega_0 k T_s) \\ \Lambda'_{q_k} := \Lambda'_m \sin(\theta'), & a'_{q_k} := \cos(\omega_0 k T_s) \\ \Lambda'_{0_k} := \Lambda'_0 e^{-kT_s/\tau'}, & a'_{0_k} := 1 \end{cases} \quad (8)$$

Four states ($\Lambda_{d_k}$, $\Lambda_{q_k}$, $\Lambda'_{d_k}$, and $\Lambda'_{q_k}$) and four coefficients ($a_{d_k}$, $a_{q_k}$, $a'_{d_k}$, and $a'_{q_k}$) are needed for representing the repeating terms in (7) and two states ($\Lambda_{0_k}$ and $\Lambda'_{0_k}$) and two coefficients ($a_{0_k}$ and $a'_{0_k}$) are required for the decaying terms.

It should be noted that in some works, such as [30], [35], the exponential function ($e^{-kT_s/\tau}$) is considered as a coefficient to estimate $\lambda_0$. But, it requires that $\tau$ to be known beforehand [36]. As the proposed method intends to minimize the need for prior knowledge, the whole decaying term ($\Lambda_0 e^{-kT_s/\tau}$) is considered as one state [37]. In addition, as the flux linkage is stable with no sharp changes, developing the states based on the flux linkage makes the formulations more reasonable. Therefore, if $\boldsymbol{a}_{m_k} = [a_{d_k}, a_{q_k}, a_{0_k}]$, $\boldsymbol{a}'_{m_k} = [a'_{d_k}, a'_{q_k}, a'_{0_k}]$, $\boldsymbol{x}_{m_k} = [\Lambda_{d_k}, \Lambda_{q_k}, \Lambda_{0_k}]^T$, and $\boldsymbol{x}'_{m_k} = [\Lambda'_{d_k}, \Lambda'_{q_k}, \Lambda'_{0_k}]^T$ then (6) can be written as

$$i_{m_k}(\Lambda_k, \Lambda'_k) \approx \beta_1(\boldsymbol{a}_{m_k}\boldsymbol{x}_{m_k}) + \beta_2(\boldsymbol{a}_{m_k}\boldsymbol{x}_{m_k})^n + \beta'_1(\boldsymbol{a}'_{m_k}\boldsymbol{x}'_{m_k}) + \beta'_2(\boldsymbol{a}'_{m_k}\boldsymbol{x}'_{m_k})^{n'}. \quad (9)$$

The expansion of the sinusoidal component given in (3) is

$$i_s(t) = i_s^p \cos(\Theta) \sin(\omega_0 k T_s) + i_s^p \sin(\Theta) \cos(\omega_0 k T_s) \quad (10)$$

and consequently, the below states and coefficients are defined:

$$\begin{cases} i_{d_k} \coloneqq i_s^p \cos(\Theta), & c_{d_k} \coloneqq \sin(\omega_0 k T_s) \\ i_{q_k} \coloneqq i_s^p \sin(\Theta), & c_{q_k} \coloneqq \cos(\omega_0 k T_s) \end{cases}. \quad (11)$$

Therefore, two states ($i_{d_k}$ and $i_{q_k}$) and two coefficients ($c_{d_k}$ and $c_{q_k}$) are required for representing the sinusoidal part of the transformer current ($i_s$). If $\boldsymbol{c}_k = [c_{d_k}, c_{q_k}]$ and $\boldsymbol{x}_{s_k} = [i_{d_k}, i_{q_k}]^T$, then

$$i_{s_k} = \boldsymbol{c}_k \boldsymbol{x}_{s_k}. \quad (12)$$

The state-space representation of current measurements is

$$\begin{cases} \boldsymbol{x}_{k+1} = \boldsymbol{x}_k + \boldsymbol{w}_k \\ \tilde{\iota}_k = h(\boldsymbol{a}_k, \boldsymbol{x}_k) + e_k \end{cases} \quad (13)$$

where the state vector is $\boldsymbol{x}_k = [\Lambda_{d_k}, \Lambda_{q_k}, \Lambda_{0_k}, i_{d_k}, i_{q_k}]^T$ for single-phase transformers or $\boldsymbol{x}_k = [\Lambda_{d_k}, \Lambda_{q_k}, \Lambda_{0_k}, \Lambda'_{d_k}, \Lambda'_{q_k}, \Lambda'_{0_k}, i_{d_k}, i_{q_k}]^T$ for core-type three-phase transformers, and the coefficient vector is $\boldsymbol{a}_k = [a_{d_k}, a_{q_k}, a_{0_k}, c_{d_k}, c_{q_k}]$ for single-phase transformers or $\boldsymbol{a}_k = [a_{d_k}, a_{q_k}, a_{0_k}, a'_{d_k}, a'_{q_k}, a'_{0_k}, c_{d_k}, c_{q_k}]$ for core-type three-phase transformers. $h(\boldsymbol{a}_k, \boldsymbol{x}_k) = i_{m_k} + i_{s_k}$ is the observation function that is the combination of (9) and (12), and $\boldsymbol{w}_k$ is the process noise vector. With reference to Fig. 2, it should be noted that one AEKF is allocated to each measured current. Therefore, $\tilde{\iota}_k$ is the $k$-th measured sample of current, and its noise is a single random variable $e_k$. As the states are considered invariant, the state-transition matrix is $\boldsymbol{I}$ (i.e., the identity matrix) [35], [37] which is unnecessary to be written in the process dynamic equation.

Each measured current sample $\tilde{\iota}_k$ is sent to AEKF with the following formulation [38],

$$\begin{cases} \boldsymbol{G}_k = \boldsymbol{P}_k^- \boldsymbol{H}_k^T (\boldsymbol{H}_k \boldsymbol{P}_k^- \boldsymbol{H}_k^T + (\hat{\sigma}_{k-1}^-)^2)^{-1} \\ \hat{\boldsymbol{x}}_k = \hat{\boldsymbol{x}}_k^- + \boldsymbol{G}_k(\tilde{\iota}_k - h(\boldsymbol{a}_k, \boldsymbol{x}_k)) \\ \boldsymbol{P}_k = (\boldsymbol{I} - \boldsymbol{G}_k \boldsymbol{H}_k)\boldsymbol{P}_k^- \\ \hat{\boldsymbol{x}}_{k+1}^- = \hat{\boldsymbol{x}}_k \\ \boldsymbol{P}_{k+1}^- = \boldsymbol{P}_k + \boldsymbol{Q} \end{cases} \quad (14)$$

where $\boldsymbol{G}$ is the gain matrix, $\boldsymbol{P}$ is the error covariance matrix, $\boldsymbol{Q}$ is the process noise covariance matrix. $(\hat{\sigma}_{k-1}^-)^2$ is the variance of the measurement noise that is estimated in AEKF as explained in "C. Noise Parameter Estimator Module". $(.)^-$ and $(.)^T$ respectively indicate prediction and matrix transposition. $\boldsymbol{H}_k = \left[\frac{\partial h}{\partial \Lambda_{d_k}}, \frac{\partial h}{\partial \Lambda_{q_k}}, \frac{\partial h}{\partial \Lambda_{0_k}}, \frac{\partial h}{\partial i_{d_k}}, \frac{\partial h}{\partial i_{q_k}}\right]$ for single-phase transformers or $\boldsymbol{H}_k = \left[\frac{\partial h}{\partial \Lambda_{d_k}}, \frac{\partial h}{\partial \Lambda_{q_k}}, \frac{\partial h}{\partial \Lambda_{0_k}}, \frac{\partial h}{\partial \Lambda'_{d_k}}, \frac{\partial h}{\partial \Lambda'_{q_k}}, \frac{\partial h}{\partial \Lambda'_{0_k}}, \frac{\partial h}{\partial i_{d_k}}, \frac{\partial h}{\partial i_{q_k}}\right]$ for core-type three-phase transformers. Then, the elements of $\boldsymbol{H}_k$ are

$$\begin{cases} \partial h/\partial \Lambda_{d_k} = a_{d_k}\left(\beta_1 + n\beta_2(\boldsymbol{a}_{m_k}\boldsymbol{x}_{m_k})^{n-1}\right) \\ \partial h/\partial \Lambda_{q_k} = a_{q_k}\left(\beta_1 + n\beta_2(\boldsymbol{a}_{m_k}\boldsymbol{x}_{m_k})^{n-1}\right) \\ \partial h/\partial \Lambda_{0_k} = \beta_1 + n\beta_2(\boldsymbol{a}_{m_k}\boldsymbol{x}_{m_k})^{n-1} \\ \partial h/\partial \Lambda'_{d_k} = a'_{d_k}\left(\beta'_1 + n'\beta'_2(\boldsymbol{a}'_{m_k}\boldsymbol{x}'_{m_k})^{n'-1}\right) \\ \partial h/\partial \Lambda'_{q_k} = a'_{q_k}\left(\beta'_1 + n'\beta'_2(\boldsymbol{a}'_{m_k}\boldsymbol{x}'_{m_k})^{n'-1}\right) \\ \partial h/\partial \Lambda'_{0_k} = \beta'_1 + n'\beta'_2(\boldsymbol{a}'_{m_k}\boldsymbol{x}'_{m_k})^{n'-1} \\ \partial h/\partial i_{d_k} = c_{d_k} \\ \partial h/\partial i_{q_k} = c_{q_k} \end{cases} \quad (15)$$

*B. Estimation Validity Evaluator Module*

It takes a while in the beginning for estimations to become stable. Therefore, a buffering time should be elapsed before sending the estimations to the next steps, as shown in Fig. 2.

An estimation is valid provided that its corresponding residual only belongs to noises [39]. Since the proposed method is real-time, the evaluation process must assess the estimated sample $k$ independently of future samples. Therefore, the hypothesis testing is employed in the proposed method. But, first, the residuals should be estimated as

$$\hat{r}_k = \tilde{\iota}_k - h(\boldsymbol{a}_k, \hat{\boldsymbol{x}}_k) \text{ and } \hat{r}_0 = 0 \quad (16)$$

where $\hat{r}_k$ is the estimated residual related to the sample $k$ and $\hat{\boldsymbol{x}}_k$ is the most updated state vector. However, normalized residuals are required for the hypothesis testing necessitating the calculation of the mean and standard deviation of the residuals. No considerable change in the noise mean and variance is expected during the time between Samples $k-1$ and $k$ unless an abrupt change occurs in the measurements. Thus, it is reasonable to use the mean and variance found for the previous Residual $k-1$ for normalizing the $k$-th residual.

The AEKF updates the measurement noise variance, recursively. Thus, the most updated standard deviation of the measurement noise is $\hat{\sigma}_{k-1} = \sqrt{(\hat{\sigma}_{k-1}^-)^2}$ that is described in "C. Noise Parameter Estimator Module". The mean of residuals is also needed for normalization that can recursively be found using

$$\begin{cases} \Psi_k = \Psi_{k-1} + \hat{r}_{k-1}, \forall k \geq 1 \\ \Psi_k = \Psi_k - \hat{r}_{k-1-m}, \forall k > m \\ \bar{r}_{k-1} = (1/m)\Psi_k, \forall k \geq m \end{cases} \quad (17)$$

where $\Psi_k$ is a variable that holds the recursive sum of the last $m$ residuals from $k-1-m$ to $k-1$ with the initial value of zero ($\Psi_0 = 0$) and $\bar{r}_{k-1}$ is the mean of $m$ residuals from $k-1-m$ to $k-1$. The $k$-th residual is then normalized as follows:

$$r_k^N = \frac{(\hat{r}_k - \bar{r}_{k-1})}{\hat{\sigma}_{k-1}}, \forall k \geq m. \quad (18)$$

The hypothesis testing is employed to evaluate if the $k$-th normalized residual belongs to the normal Gaussian random variables (i.e., $\mathcal{N}(0,1)$) with a probability of false alarms of $\rho$ ($0 < \rho < 1$) for which a small value is preferred [39]. Using the Probability Distribution Function (PDF) of the standard Gaussian random variable, the threshold $T$ is found as [39]

$$0.5 - \int_0^T p(\xi)d\xi = \rho/2 \quad (19)$$

where $T$ is the threshold, $\xi$ is the integration variable, and $p(.)$ is the PDF of the normal Gaussian random variable. Further explanations are provided in Appendix B. If the normalized residual is smaller than the threshold $T$, the estimation is valid, otherwise it is invalid as [39]

$$\begin{cases} |r_k^N| < T \rightarrow Flag = 0 \\ |r_k^N| \geq T \rightarrow Flag = 1 \end{cases}. \quad (20)$$

It should be noted that the value of $T$ is selected one time and plugged in (20). Therefore, the integration in (19) does not add to the computational burden of the proposed method.

*C. Noise Parameter Estimator Module*

The AEKF estimates the measurement noise variance as [38],

$$\begin{cases} (\hat{\sigma}_{k-1}^-)^2 = \sigma_0^2, \forall k \leq m \\ (\hat{\sigma}_{k-1}^-)^2 = \frac{1}{m}\sum_{i=1}^m \hat{r}_{k-1-i}^2 - \boldsymbol{H}_{k-1}\boldsymbol{P}_{k-1}^-\boldsymbol{H}_{k-1}^T, \forall k > m \end{cases} \quad (21)$$

where $\sigma_0^2$ is the initial measurement noise variance, $m$ is the number of past residuals for updating the measurement noise. The summation term in (21) is computationally expensive. Therefore, (22) is proposed to be used instead of (21) for higher computational performance.

$$\begin{cases} (\hat{\sigma}_{k-1}^-)^2 = \sigma_0^2, \forall k < m \\ \psi_k = \psi_{k-1} + \hat{r}_{k-1}^2, \forall k \geq 1 \\ \psi_k = \psi_k - \hat{r}_{k-1-m}^2, \forall k > m \\ (\hat{\sigma}_{k-1}^-)^2 = (1/m)\psi_k - \boldsymbol{H}_{k-1}\boldsymbol{P}_{k-1}^-\boldsymbol{H}_{k-1}^T, \forall k \geq m \end{cases} \quad (22)$$

where $\psi_k$ recursively retains the sum of the squares of the past $m$ residuals. As a result, $m-1$ "summation operations" in (21) are replaced with two summation operations, technically. This amendment improves the computational performance of the proposed method considering that $m$ is a relatively large number. It is also worth noting that the "division operation" is computationally more expensive compared to the "multiplication operation". Therefore, it is suggested that the value of $(1/m)$ in (17) and (22) is calculated offline and plugged into those formulas as a constant multiplier.

*D. Interpolator Module*

The sampling frequency of commonplace measuring devices is relatively low (e.g., 2,880 [40] and 30,720 [41]). However, suitable time steps for replicating transformer transients are in the range of some microseconds. Therefore, the sampling rates of measurements should be increased as follows,

$$\text{Flag} = 0 \rightarrow \hat{\imath}_j = h(\boldsymbol{a}_j, \hat{\boldsymbol{x}}_{k-1}) \quad (23)$$

where $\hat{\imath}_j$ is the reconstructed sample $j$ at a high sampling rate suitable for the solver, in the case of single-phase transformers,

$$\boldsymbol{a}_j = [\sin(\omega_0 j\Delta t), \cos(\omega_0 j\Delta t), 1, \sin(\omega_0 j\Delta t), \\ \cos(\omega_0 j\Delta t)] \quad (24)$$

and for core-type three-phase transformers,

$$\boldsymbol{a}_i = [\sin(\omega_0 j\Delta t), \cos(\omega_0 j\Delta t), 1, \sin(\omega_0 j\Delta t), \\ \cos(\omega_0 j\Delta t), 1, \sin(\omega_0 j\Delta t), \cos(\omega_0 j\Delta t)] \quad (25)$$

where $\Delta t$ is the solver time step. However, if the estimation is recognized as invalid (i.e., Flag = 1), an abrupt change probably occurred and proper measures should be taken which are beyond the scope of this paper.

It is suggested that the values of $\sin(.)$ and $\cos(.)$ functions are calculated offline, and the results are saved in lookup tables. Then, their values are fetched from lookup tables to enhance the computational effectiveness of the proposed method.

III. TEST CASES AND DISCUSSION

One single-phase transformer and one three-limb, core-type, three-phase transformer are simulated in EMTP-RV utilizing the trapezoidal rule of integration with a time step of 2 μs. The simulation results are then downsampled at a ratio of 100 to 1 to make the sampling frequency ($f_s$) of the measured currents 5 kHz. This low sampling frequency lessens the trackability of the proposed method, reducing its performance. Moreover, a zero-mean Gaussian noise with a standard deviation equal to three percent of the transformer nominal current (i.e., $\mathcal{N}(0, (0.03 i_{nom})^2)$) is added to the measurement samples. The probability of false alarms is chosen as $\rho = 0.01$ with a corresponding threshold $T = 2.575$ (i.e., $0.5 - \int_0^T p(\xi)d\xi = 0.01/2$ and the value of $T$ is found 2.575 from one-tailed standard Gaussian tables). The window length for calculation of the residual mean and noise variance ($m$ in (17) and (22)) is 100.

*A. Single-Phase Transformer*

The single-phase transformer is 380:220 V, 5 kVA, and 60 Hz with parameters given in Table I from [22]. Its model (shown in Fig. 3(a)) is developed in EMTP-RV for running tests. In the first test, the transformer is energized from the LV side (i.e., X side in Fig. 3(a)) at its nominal voltage while its HV side remains open. The measured and estimated steady-state no-load currents are depicted in Fig. 4(a). It is obvious that the estimated current, which is found with $h(\boldsymbol{a}_k, \hat{\boldsymbol{x}}_k)$, is valid although the noises are considerable. The largest difference between the estimated and measured values is encircled in Fig. 4(a) at $t = 10.414$ s. The corresponding residuals resulting from the differences between measurements and estimated values are calculated using (16) and shown in Fig. 4(b). The residual at $t = 10.414$ s is the largest residual, and it is highlighted with a red circle. Fig. 4(c) shows the normalized residuals ($r_k^N$) in the same time window. The normalized residual caused by the largest residual at $t = 10.414$ s exceeds the threshold of 2.575, and it is flagged in Fig. 4(c).

TABLE I
PARAMETERS OF THE SINGLE-PHASE TRANSFORMER [22]

| $R_H = 0.43\ \Omega$ | $R_X = 0.14\ \Omega$ | $L_H = 810\ \mu H$ | $L_X = 270\ \mu H$ |
|---|---|---|---|
| $R_C = 1310\ \Omega$ | $\beta_1 = 0.054$ | $\beta_2 = 0.039$ | $n = 5$ |

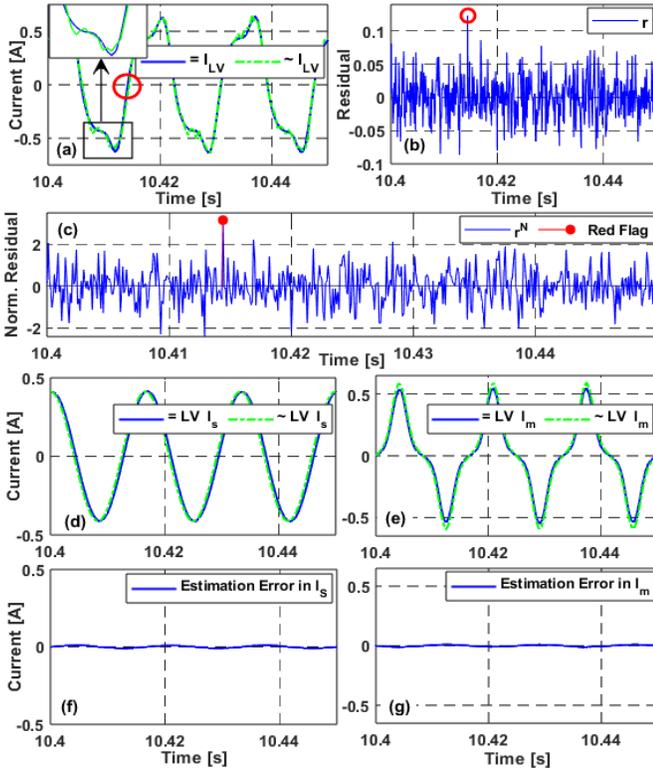

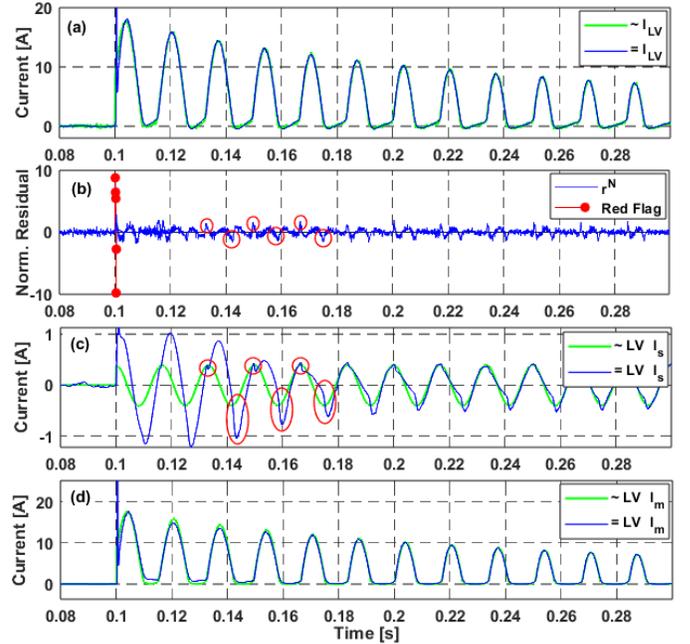

Fig. 4. No-load test. "=" and "~" respectively indicate estimated and measured values. (a) Measured and estimated no-load currents on the LV side. (b) The residuals of current estimation. (c) The normalized residuals. (d) and (e) Estimated and the actual values of the magnetization current component. (f) and (g) The absolute errors in estimations of sinusoidal and magnetization components.

Fig. 4(d) shows the estimated sinusoidal component of the transformer current ($i_s$) that passes through $R_c$ (in Fig. 3(a)) in no-load conditions. Although it is impossible to directly measure the $R_c$'s current in practice, it is measurable in the simulation and shown in Fig. 4(d) with "~ LV $I_s$". It is visually clear that the estimations and measurements are very close. Fig. 4(e) depicts the estimated and measured magnetization currents ($i_m$). Again, the magnetization current cannot be directly measured in practice. However, it can be measured in the simulation. Figs. 4(f) and (g) show the error in the estimation of sinusoidal and magnetization components, respectively. Although a pattern is recognized in both errors, it is acceptable considering its negligible size, the complexity of the model, the approximations in relating total flux linkage to current in (2) and (6), and the fact that no noise is added to the measured sinusoidal and magnetization components as those measurements are only available in the simulation and practically unavailable.

Fig. 5 illustrates the performance of the proposed method in the case of inrush currents in that the LV side is energized at $t = 0.1$ s, while the HV side remains open. The initial flux of the core is zero. The LV-side current is considered since it is larger than that of the HV side and makes the situation more challenging for the proposed method. Fig. 5(a) depicts the measured and estimated inrush currents on the LV side of the transformer. It is obvious that the estimation becomes valid fast although deep saturation occurs. Fig. 5(b) shows the normalized residuals related to the estimation of the inrush current. Some small peaks in the normalized residuals are encircled that are caused by decreasing mismatches between the estimated and measured sinusoidal components ($i_s$) which are encircled in Fig. 5(c). The mentioned peaks are much smaller than the threshold ($T = 2.575$) and gradually disappear with no discernable error in the total estimated transformer current. In contrast to the sinusoidal component $i_s$, the estimated magnetization component matches the measured one shortly after switching as Fig. 5(d) shows.

Fig. 5. Energization test. "~" and "=" respectively indicate measured and estimated values. (a) Measured and estimated inrush currents. (b) Normalized residuals related to the transformer current. (c) Measured and estimated sinusoidal components. (d) Measured and estimated magnetization components.

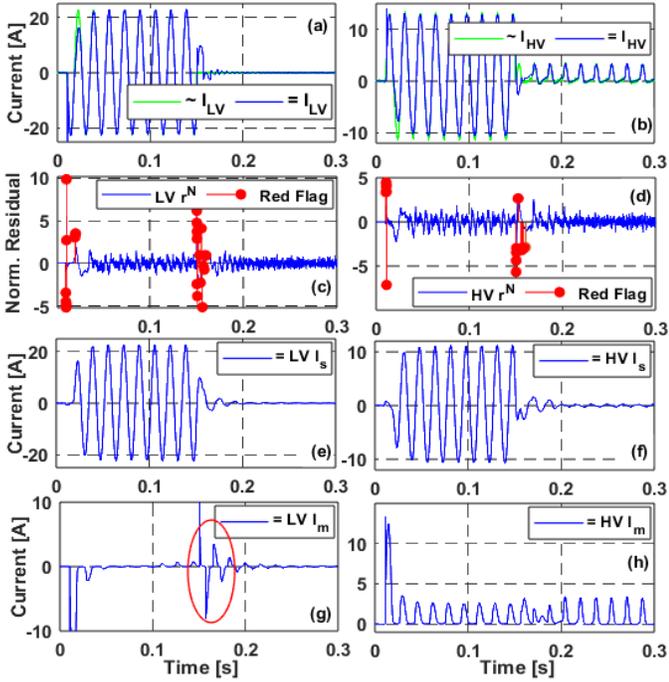

**Fig. 6.** Underload switching. "~" and "=" respectively indicate measured and estimated values. (a) and (b) Measured and estimated LV and HV currents. (c) and (d) Normalized residuals corresponding to LV and HV sides. (e) and (f) Estimated sinusoidal components in LV- and HV-side currents. (g) and (h) Estimated magnetization components in LV- and HV-side currents.

TABLE II
MSEs FOR DIFFERENT SWITCHING ANGLES

| $\alpha°$ | 0 | 10 | 20 | 30 | 40 |
|---|---|---|---|---|---|
| MSE | 0.166 | 0.133 | 0.098 | 0.101 | 0.090 |
| $\alpha°$ | 50 | 60 | 70 | 80 | 90 |
| MSE | 0.082 | 0.115 | 0.153 | 0.167 | 0.193 |

TABLE III
MSEs FOR DIFFERENT TRANSFORMER LOADING

| LV voltage | $219\angle-0.5°$ | $218.5\angle-0.5°$ | $218\angle-0.5°$ |
|---|---|---|---|
| MSE | 0.466 | 0.443 | 0.421 |
| LV voltage | $219\angle-1°$ | $219\angle-1.5°$ | $219\angle-2°$ |
| MSE | 0.323 | 0.217 | 0.205 |

TABLE IV
PARAMETERS OF THE THREE-PHASE TRANSFORMER [42]

| P* | $R_H$ | $R_X$ | $R_y$ | $R_1$ | $R_4$ | $X_3$ | $X_4$ |
|---|---|---|---|---|---|---|---|
| A | 0.1754 | 0.0393 | 312.74 | 412.88 | 0.2112 | 0.0267 | 0.0370 |
| B | 0.1786 | 0.0389 | 312.74 | 412.88 | 0.2112 | 0.0273 | 0.0370 |
| C | 0.1771 | 0.0389 | 312.74 | 412.88 | 0.2112 | 0.0277 | 0.0370 |

* P stands for Phase.

The inrush test is repeated for different switching angles ($\alpha$). The Mean Squared Errors (MSEs) (i.e., $MSE = \frac{1}{500}\sum_{k=1}^{500}(r_k^N)^2$) for 0.1 s (i.e., 500 samples) after switchings are found for each switching angle and presented in Table II. It is noticed that the MSEs corresponding to switching angles near 45° are slightly smaller, meaning that the estimations include less error when the switching angle is close to 45°. It is also noticed that the estimated magnetization component becomes valid considerably faster than the sinusoidal component regardless of the switching angle.

The performance of the proposed method in underload switching of the transformer is shown in Fig. 6. The HV and LV sides of the transformer are respectively switched to sources with the voltages of $380\angle 0°$ and $219\angle -1.5°$ V at $t = 0.01$ s. The differences in voltage magnitudes and phases cause a flow of reactive and active powers through the transformer. The LV side becomes disconnected at $t = 0.15$ s, but the transformer stays energized from the HV side.

Figs. 6(a) and (b) show the measured and estimated LV and HV currents. The transient parts of the currents decay fast as the transformer is under load. However, the proposed method is capable of tracking the currents. It is also clear that the LV current becomes zero after 0.15 s when it becomes disconnected from the LV side. However, the HV side stays connected and draws the magnetization current.

As Figs 6(c) and (d) depict, the red flags are released after the connection of both sides at $t = 0.01$ s and after the disconnection of the LV side at $t = 0.15$ s. The red flags are released on both sides due to the abrupt changes in both LV and HV currents due to the LV side switching. For evaluation of estimation errors in underload switching, the MSE of the normalized residuals for 0.1 s after the moment of connection is calculated for different transformer loadings. Different loadings are made by keeping the HV side voltage constant and equal to $380\angle 0°$ V, while the magnitude and phase of the LV side voltage vary for passing different reactive and active powers through the transformer. The resultant MSEs are given in Table III. It is noticed that lower MSEs (indicating less estimation errors) occur when a higher portion of the power is active (i.e., larger phase difference) as the shaded cell in the table indicates.

Figs. 6(e) and (f) show the estimated sinusoidal components. It is found that when the sinusoidal current is comparable to or larger than the magnetization current, it is correctly estimated in a shorter time compared to the no-load situation. Figs. 6(g) and (h) show the estimated magnetization components. It is clear that a large fluctuation occurs in the estimation when a current is suddenly cut as encircled in Fig. 6(g).

*B. Core-Type Three-Phase Transformer*

The hybrid model (Fig. 3(b)) of a core-type, three-limb, 15-kVA, 60-Hz, 208Y/240Δ distribution transformer is made in EMTP-RV with the given parameters in Table IV from [42].

The transformer is energized with its nominal voltage at 0.016 s from the LV side, while its HV side remains open. The initial fluxes of all inductors are zero. As energization from the LV side results in larger inrush currents, it is selected for a decent evaluation of the proposed method.

Figs. 7(a), (b), and (c) illustrate the measured and estimated currents of Phases A, B, and C, respectively. Although the core becomes highly saturated, the proposed method is able to estimate the currents since the normalized residuals shown in Figs. 7(d), (e), and (f) mostly consist of noises after a short

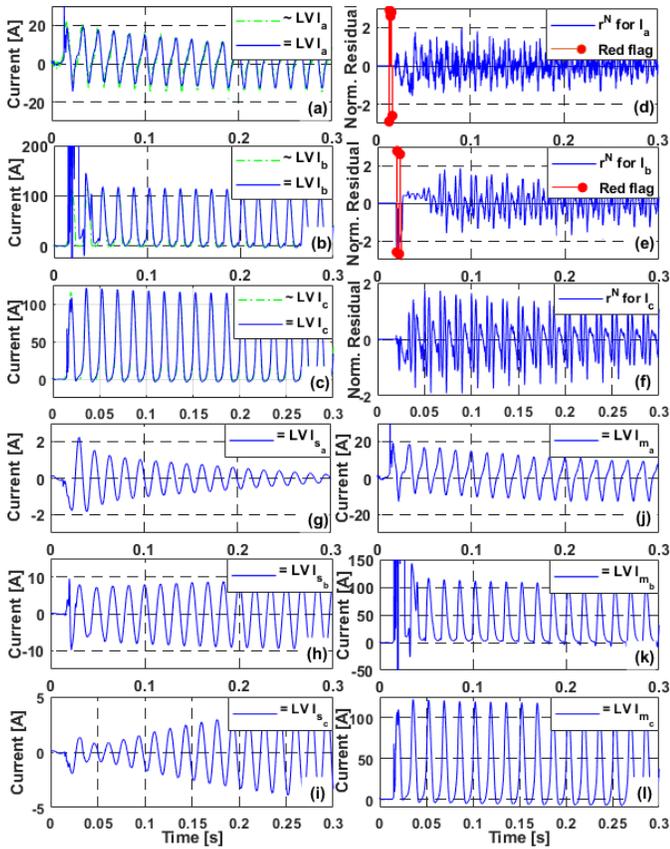

**Fig. 7.** Energization test. "~" and "=" respectively indicate measured and estimated values. (a), (b), and (c) Measured and estimated currents of Phases A, B and C. (d), (e) and (f) Normalized residuals. (g), (h), and (i) Estimated sinusoidal components. (j), (k), and (l) Estimated magnetization components.

while. No invalid output is detected in estimating the Phase C current as there is no red flag in Fig. 7(f). However, red flags are highly probable in the beginning for other switching times or initial values of states. Although some patterns can be recognized in the normalized residuals, those are insignificant in light of that the patterns are negligible and the complex behavior of transformer cores results in approximate models causing some mismatches between the measurements and the observer equation.

Figs. 7(g), (h), and (i) respectively show the estimated sinusoidal components in Phases A, B, and C currents. As the sinusoidal component is related to several components of the transformer, it is impossible to measure its actual value for comparison purposes. The magnetization components of Phases A, B, and C currents are respectively shown in Figs. 7(j), (k), and (l). Neglecting the predictable invalid estimations in the beginning, their estimated values become valid fast. Again, as the magnetization currents are caused by different yoke and limb nonlinear inductors (Fig. 3(b)), it is impossible to measure and compare them with the estimated values. However, the normalized residuals in Figs. 7(d), (e), and (f) indicate the combination of errors in the estimated sinusoidal and magnetization components. As the normalized residuals mainly contain noises, it can be concluded that both sinusoidal and magnetization parts are successfully estimated.

The overall effect of the proposed method in improving the performance of DTs is discussed with reference to Fig. 8. The measured no-load current of Phase A without additive noises with $f_S = 5$ kHz is shown in Fig. 8(a). Fig. 8(b) shows the same measurement but with additive noises. Fig. 8(c) depicts the reconstructed measurement with a sampling rate of 500 kHz, and Fig. 8(d) shows the simulated current as the reference with $f_S = 500$ kHz. Using the trapezoidal rule of integration (i.e., $area = (\Delta t/2) \sum \hat{\imath}_{i+1} + \hat{\imath}_i$) the areas between the measurements and the time axis are calculated and given in Table V. The relative errors in calculation of the areas with respect to the reference (Fig. 8(d)) are also given in Table V. The relative errors are exceedingly large without the application of the proposed method as the errors in the first two columns indicate. Conversely, the error associated with the reconstructed measurement is slight.

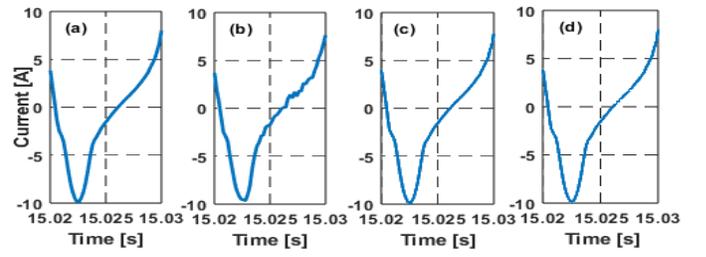

**Fig. 8.** (a) Low-sampling rate, but noiseless measurement. (b) Low sampling rate and noise-contaminated measurement. (c) Reconstructed measurement at a high sampling rate. (d) Simulated measurement as the reference case.

TABLE V
The Effect of Measurement Reconstruction

| $f_S$ | Measured Noiseless | Measured Noisy | Reconstructed | Reference |
|---|---|---|---|---|
| Area | -0.011618 | -0.011273 | -0.013372 | -0.013443 |
| Relative Error % | 13.58 | 16.14 | 0.53 | 0 |

## CONCLUSION

In this paper, a recursive method for the development of measurement integrity verification suitable for transformer currents is proposed. The Hardware-in-the-Loop (HIL) Digital Twins (DTs) replicating transformers require numerically solving the Differential Equations (DEs) governing transformer transients with small time steps in the order of microseconds. Electrical measurements are often noise-contaminated and low sampling. However, the flow of noiseless and high-rate measurements to the DT is crucial. Accordingly, the proposed measurement integrity verification ensures the quality of current measurement flows. In addition, the proposed method, owing to its modular structure, can be utilized in gradual extension of DTs.

The Adaptive Extended Kalman Filter (AEKF) is employed in this work to estimate the parameters of measured currents. Then the estimations are evaluated using the hypothesis test, and the valid ones are used in the reconstruction of measurements at the desired sampling rate. The performance of the proposed method is evaluated using EMTP-RV for running simulations and Matlab for the execution of the proposed method. Several different scenarios are defined, and the results demonstrate that the proposed method is capable of satisfactorily estimating the current measurement parameters and tracking the changes in transformer currents even if the transformer core is intensely saturated.

## APPENDIX A

The magnetization current ($i_m$) of each phase can be divided into two parts as

$$i_m = i_L + i_Y \tag{A.1}$$

where $i_L$ and $i_Y$ are parts of the magnetization current that are respectively induced by limb and yoke inductors. However, there are different contributions from inductors in $i_L$ and $i_Y$ as

$$\begin{aligned} i_L &= k_1 i_{L_1} + k_2 i_{L_2} + k_3 i_{L_3} \\ i_Y &= k_4 i_{Y_1} + k_5 i_{Y_2} \end{aligned} \tag{A.2}$$

where $k_1, k_2, \ldots, k_5$ are scalars between zero and one representing the different contributions of core inductor currents (limb and yoke) in $i_L$ and $i_Y$. Therefore, $i_L$ is a linear combination of nonlinear currents passing through limb inductors ($L_{1_1}$, $L_{1_2}$, and $L_{1_3}$), and $i_Y$ is also a linear combination of the nonlinear currents in yoke inductors ($L_{y_1}$ and $L_{y_2}$).

**Remark 1**: Based on the closure under addition and scalar multiplication property, the linear combinations of functions of a certain type (e.g., polynomials, trigonometric functions) results in a function belonging to the same type [43].

Therefore, the general shape of $i_L$ and $i_Y$ will be the same as their constituent currents.

## APPENDIX B

In the hypothesis testing, two statements should be defined. The null hypothesis ($H_0$) that is the statement to be tested, and its complement which is called the alternative hypothesis ($H_1$) [39]. $H_0$ in this paper states that the normalized residual $k$ ($r_k^N$) is caused by noises, and $H_1$ is that $r_k^N$ is not caused by noises. A false alarm is released when the residual is induced by noises, but it is found to be caused by other events.

For $r_k^N$ to belong to the noise $\mathcal{N}(0,1)$ with the confidence level of P, $r_k^N$ must be between $-T$ and $T$ ($|r_k^N| \leq T$) where $\int_{-T}^{T} p(\xi)d\xi = P$ where $p(.)$ is the PDF of the standard Gaussian random variable. The complement of this probability ($H_1$) is $|r_k^N| > T$, and $1 - \int_{-T}^{T} p(\xi)d\xi = 1 - P$. However, as $r_k^N$ is a positive-valued number, one-tailed integration of $p(.)$ is enough which is

$$0.5 - \int_0^T p(\xi)d\xi = \frac{(1-P)}{2} \tag{B.1}$$

where $1 - P$ is the complement of P which is shown with $\rho$ that means the probability of false alarms, meaning that $r_k^N$ belongs to the noise but recognized as the effect of other events.

> REPLACE THIS LINE WITH YOUR MANUSCRIPT ID NUMBER (DOUBLE-CLICK HERE TO EDIT) <

10twin technology," *IEEJ Trans. Electr. Electron. Eng.*, vol. 17, no. 11, pp. 1629–1636, Nov. 2022, doi: 10.1002/tee.23670.

[20] R. Jalilzadeh Hamidi, "Protective relay for power transformers based on digital twin systems," in *2022 IEEE Kansas Power and Energy Conference (KPEC)*, Manhattan, KS, Apr. 2022, pp. 1–6. doi: 10.1109/KPEC54747.2022.9814786.

[21] S. Prusty and M. Rao, "A direct piecewise linearized approach to convert rms saturation characteristic to instantaneous saturation curve," *IEEE Trans. Magn.*, vol. 16, no. 1, pp. 156–160, Jan. 1980, doi: 10.1109/TMAG.1980.1060567.

[22] H. Dirik, C. Gezegin, and M. Özdemir, "A novel parameter identification method for single-phase transformers by using real-time data," *IEEE Trans. Power Deliv.*, vol. 29, no. 3, pp. 1074–1082, Jun. 2014, doi: 10.1109/TPWRD.2013.2284243.

[23] IEEE Standard Definitions for the Measurement of Electric Power Quantities Under Sinusoidal, Nonsinusoidal, Balanced, or Unbalanced Conditions, IEEE Std 1459-2000, Mar. 2010, doi: 10.1109/IEEESTD.2010.5439063.

[24] A. A. Girgis, W. B. Chang, and E. B. Makram, "A digital recursive measurement scheme for online tracking of power system harmonics," *IEEE Trans. Power Deliv.*, vol. 6, no. 3, pp. 1153–1160, Jul. 1991, doi: 10.1109/61.85861.

[25] S. Kashyap and A. K. Singh, "A comparative study of WPT and DWT based techniques for measurement of harmonics," in *2008 13th Int. Conf. on Harmonics and Quality of Power*, Wollongong, NSW, Australia, Oct. 2008, pp. 1–5. doi: 10.1109/ICHQP.2008.4668869.

[26] K. Hasan, M. M. Othman, S. T. Meraj, S. Mekhilef, and A. F. B. Abidin, "Shunt active power filter based on Savitzky-Golay filter: Pragmatic modelling and performance validation," *IEEE Trans. Power Electron.*, vol. 38, no. 7, pp. 8838–8850, Jul. 2023, doi: 10.1109/TPEL.2023.3258457.

[27] N. A. Yalcin and F. Vatansever, "A new hybrid method for signal estimation based on Haar transform and Prony analysis," *IEEE Trans. Instrum. Meas.*, vol. 70, pp. 1–9, Sep. 2020, doi: 10.1109/TIM.2020.3024358.

[28] L. Bernard, S. Goondram, B. Bahrani, A. A. Pantelous, and R. Razzaghi, "Harmonic and interharmonic phasor estimation using matrix pencil method for phasor measurement units," *IEEE Sens. J.*, vol. 21, no. 2, pp. 945–954, Jan. 2021, doi: 10.1109/JSEN.2020.3009643.

[29] R. Jalilzadeh Hamidi, H. Livani, and R. Rezaiesarlak, "Traveling-wave detection technique using short-time matrix pencil method," *IEEE Trans. Power Deliv.*, vol. 32, no. 6, pp. 2565–2574, Dec. 2017, doi: 10.1109/TPWRD.2017.2685360.

[30] H. C. Wood, N. G. Johnson, and M. S. Sachdev, "Kalman filtering applied to power system measurements relaying," *IEEE Trans. Power Appar. Syst.*, vol. PAS-104, no. 12, pp. 3565–3573, Dec. 1985, doi: 10.1109/TPAS.1985.318911.

[31] A. Akrami and H. Mohsenian-Rad, "Event-triggered distribution system state estimation: sparse Kalman filtering with reinforced coupling," *IEEE Trans. Smart Grid*, vol. 15, no. 1, pp. 627–640, Jan. 2024, doi: 10.1109/TSG.2023.3270421.

[32] IEEE Standard Requirements for Power-Line Carrier Line Traps (30 kHz to 500 kHz), IEEE Std C93.3-2017, Feb. 2017. doi: 10.1109/IEEESTD.2017.8000798.

[33] P. Han *et al.*, "An inrush current suppression strategy for UHV converter transformer based on simulation of magnetic bias," *IEEE Trans. Power Deliv.*, vol. 37, no. 6, pp. 5179–5189, Dec. 2022, doi: 10.1109/TPWRD.2022.3173431.

[34] J. Mitra, X. Xu, and M. Benidris, "Reduction of three-phase transformer inrush currents using controlled switching," *IEEE Trans. Ind. Appl.*, vol. 56, no. 1, pp. 890–897, Feb. 2020, doi: 10.1109/TIA.2019.2955627.

[35] D. D. Patel, K. D. Mistry, M. B. Raichura, and N. G. Chothani, "Three state Kalman filter based directional protection of power transformer," in *2018 20th National Power Systems Conference (NPSC)*, Tiruchirappalli, India, Dec. 2018, pp. 1–6. doi: 10.1109/NPSC.2018.8771716.

[36] F. B. Ajaei and M. Sanaye-Pasand, "Minimizing the impact of transients of capacitive voltage transformers on distance relay," in *2008 Int. Conf. on Pow. Sys. Tech. and IEEE Pow. India Conf.*, Delhi, India, Oct. 2008, pp. 1–6. doi: 10.1109/ICPST.2008.4745229.

[37] A. Hooshyar and M. Sanaye-Pasand, "Accurate measurement of fault currents contaminated with decaying DC offset and CT saturation," *IEEE Trans. Power Deliv.*, vol. 27, no. 2, pp. 773–783, Apr. 2012.

[38] R. Mehra, "Approaches to adaptive filtering," *IEEE Trans. Autom. Control*, vol. 17, no. 5, pp. 693–698, Oct. 1972, doi: 10.1109/TAC.1972.1100100.

[39] A. Abur and A. G. Expósito, *Power System State Estimation: Theory and Implementation*, 1st ed. Boca Raton, Fl, USA: CRC Press, 2004.

[40] *Distributed busbar protection REB500 including line and transformer protection*, ABB Switzerland Ltd, Switzerland, 1MRB520308-Ben, 2011. [Online]. Available: https://library.e.abb.com/public/ad4143d41cea2c16c1257a70001f0ed0/1MRB520308-BEN_D_en_Distributed_busbar_protection_REB500.pdf

[41] *SEL-735 Power Quality and Revenue Meter*, Schweitzer Engineering Laboratories, Inc (SEL), Pullman, WA, U.S.A., Nov. 2022. [Online]. Available: https://selinc.com/products/735/

[42] J. Mitra, X. Xu, and M. Benidris, "A controlled switching approach to reduction of three-phase transformer inrush currents," in *2018 IEEE Industry Applications Society Annual Meeting (IAS)*, Sep. 2018, pp. 1–7.

[43] L. Hogben, Ed., *Handbook of linear algebra*. in Discrete mathematics and its applications. Boca Raton, Fla.: Chapman & Hall/CRC, 2007.